\documentclass[a4paper]{spie}  \usepackage[]{graphicx}
\usepackage[numbers]{natbib}

\title{Geometrical model fitting for interferometric data: GEM-FIND } 

\author{Daniela Klotz\supit{a}, Stephane Sacuto\supit{b}, Claudia Paladini\supit{a}, Josef Hron\supit{a} and Georg Wachter\supit{c}
\skiplinehalf
\supit{a}Department of Astrophysics, University of Vienna,
              T\"urkenschanzstrasse 17, A-1180 Vienna, Austria; \\
\supit{b}Department of Physics and Astronomy, Division of Astronomy and Space Physics, Uppsala University, Box 516, 75120, Uppsala, Sweden;\\
\supit{c}Institute for Theoretical Physics, Vienna University of Technology, Wiedner Hauptstr. 8-10, A-1040 Vienna, Austria, EU;
}

\authorinfo{Further author information: Send correspondence to daniela.klotz@univie.ac.at}
 
\begin{document} 
  \maketitle 


\begin{abstract}
We developed the tool GEM-FIND that allows to constrain the morphology and brightness distribution of objects. The software fits geometrical models to spectrally dispersed interferometric visibility measurements in the $N$-band using the Levenberg-Marquardt minimization method. Each geometrical model describes the brightness distribution of the object in the Fourier space using a set of wavelength-independent and/or wavelength-dependent parameters. 
In this contribution we numerically analyze the stability of our nonlinear fitting approach by applying it to sets of synthetic visibilities with statistically applied errors, answering the following questions: How stable is the parameter determination with respect to (i) the number of $uv$-points, (ii) the distribution of points in the $uv$-plane, (iii) the noise level of the observations? 
\end{abstract}


\section{Introduction: GEM-FIND} 
\begin{figure}
\centering
\includegraphics*[width=15cm]{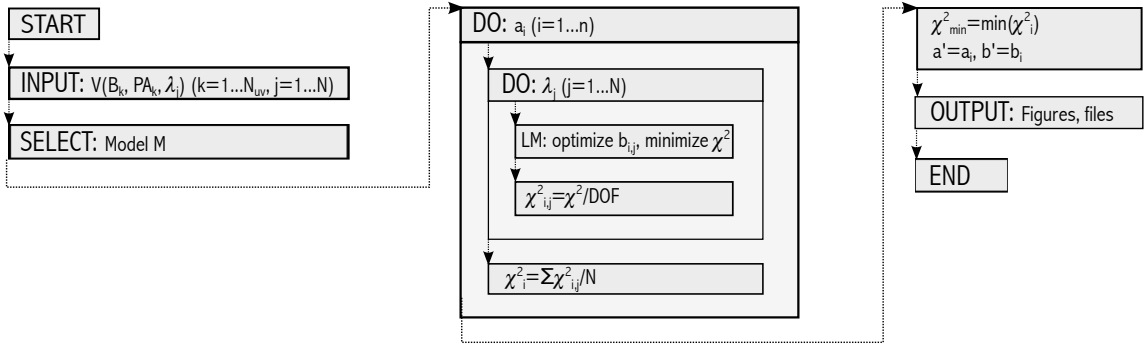}
\caption{\label{flow-chart} Flow-chart illustrating the different stages in the fitting approach. \textbf{Column A:} The observed visibilities $V$ are passed on to GEM-FIND and one of the geometrical models M1-M13 (see Table\,\ref{modelparam}) is selected. \textbf{Column B:} The wavelength-independent parameters $a_i$ are varied over an equidistant grid with $n$ grid points. A Levenberg-Marquardt (LM) least squares fitting minimization is performed to the observed visibilities for each wavelength point $\lambda_j$, i.e. the wavelenght-dependent parameters $b_{i,j}$ are optimized to minimize $\chi^2$. The best solution results in a reduced $\chi^2_{i,j}$. A color-reduced $\chi^2_i$ is derived for each grid point. \textbf{Column C:} The best-fitting wavelength-dependent and independent parameters b' and a' are determined by the grid point $i$ having the minimal $\chi_{\rm min}^2$. 
}
\end{figure}

In order to fit observations obtained with the mid-IR interferometric instrument VLTI/MIDI with geometrical models the software GEM-FIND (GEometrical Model Fitting for INterferometric Data) was developed. It fits wavelength-dispersed visibility measurements (from 8-13\,$\mu$m) with synthetic visibilities from centro-symmetric and asymmetric geometrical models. Each model represents the Fourier transformed brightness distribution of the object that is derived by means of a set of wavelength-independent and/or wavelength-dependent parameters.
The wavelength-independent parameters are varied using an equidistant grid. For each grid point a non-linear least squares fitting minimization (based on the Levenberg-Marquardt method: see Ref.\,\citenum{markwardt09}) is performed for each wavelength point on the wavelength-dependent parameters. This determines a color-reduced $\chi^2$ using
\begin{equation}
\chi_i^2=\frac{\sum\limits^j\chi_{i,j}^2}{N} \qquad\mbox{for}\;i=1\dots n, \;j=1\dots N,
\end{equation} 
\begin{table}[]
\begin{footnotesize}
\begin{center}
\caption{\label{modelparam}Parametric description of the geometrical models included in GEM-FIND.}
\begin{tabular}{llllll}
\hline\hline
\#$_\mathrm{model}$&	Model   \rule{0pt}{2.6ex}	&	\multicolumn{2}{l}{$\lambda$ independent}					&	$\lambda$ dependent						&	Application\\
				&						& fixed 				& grid								&	free										&\\
\hline
M1				&	Circular UD			&					&									&  	$\theta$						& Optically thick circular CSE\\
M2				&	Circular Gaussian		&					&									&	FWHM                     							& Optically thick circular CSE\\
M3				&	Elliptical UD			&					&$\psi$, $\eta$							&	$\theta_{\mathrm{maj}}$						& Optically thick elliptical CSE\\
M4				&	Elliptical Gaussian		&					&$\psi$, $\eta$							&	FWHM$_{\mathrm{maj}}$						& Optically thick elliptical CSE\\
M5				&	CircUD+CircGauss		&$\theta_{\mathrm{cen}}$	&									&	FWHM, $f$								& Star + optically thin circular CSE\\
M6				&	CircUD+EllGauss		&$\theta_{\mathrm{cen}}$	&$\psi$, $\eta$							&	FWHM$_{\mathrm{maj}}$, $f$					& Star + optically thin elliptical CSE\\
M7				&	UD+Dirac				&					&$\Delta x$, $\Delta y$				&	$f$, $\theta_{\mathrm{prim}}$ 					& Partially resolved binary system \\
M8				&	Dirac+Dirac			&					&$\Delta x$, $\Delta y$				&	$f$										& Unresolved binary system\\
M9				&	UD+UD				&					&$\Delta x$, $\Delta y$				&$f$, $\theta_{\mathrm{prim}}$, $\theta_{\mathrm{sec}}$	& Resolved binary system\\
M10				&	UD-UD				&					&									&	$\theta_{\mathrm{out}}$, $\theta_{\mathrm{in}}$	& Ring\\
M11				&	UD+Ring				&$\theta_{\mathrm{cen}}$	&									&	$\theta_{\mathrm{out}}$, $\theta_{\mathrm{in}}$, $f$	& Detached shell object\\
M12				&	CircGauss+UD+Dirac		&$\theta_{\mathrm{cen}}$	&$\Delta x$, $\Delta y$	&	FWHM$_{\mathrm{maj}}$, $f_1$, $f_2$			& Optically thin circular CSE + \\
				&						&					&									&										&star + companion\\
M13				&	EllGauss+UD+Dirac		&$\theta_{\mathrm{cen}}$	&$\psi$, $\eta$, $\Delta x$, $\Delta y$	&	FWHM$_{\mathrm{maj}}$, $f_1$, $f_2$			& Optically thin elliptical CSE + \\
				&						&					&									&										&star + companion\\
\hline
\end{tabular}
\end{center}
\footnotesize{\textbf{Notes.} UD\ldots uniform disk; CSE\ldots circumstellar environment; $\theta_{(maj)}$\ldots uniform disk diameter (major axis); FWHM$_{(maj)}$\ldots full width at half maximum of Gaussian (major axis); $\psi$\ldots inclination angle of ellipse; $\eta$\ldots axis ratio minor/major axis of ellipse; $\Delta x$, $\Delta y$\ldots angular offsets of binary component from center of symmetry; $f$\ldots flux ratio binary/primary or central star/envelope; $\theta_{cen}$\ldots diameter of central star; $\theta_{\mathrm{prim}}$, $\theta_{\mathrm{sec}}$\ldots diameter of primary or secondary component; $\theta_{(out)},\theta_{(in)}$\ldots outer or inner ring diameter.}
\end{footnotesize}
\end{table}
at grid point $i$ and wavelength index $j$. Note that $\chi_{i,j}^2$ is already reduced for the number of degrees of freedom for each wavelength point. The number of degrees of freedom in the fitting is solely determined by the wavelength-dependent parameters. The best-fitting parameters are then determined by 
\begin{equation}
\chi_{\rm min}^2=\min\left(\chi^2_i\right)\,. 
\end{equation}
Uncertainties of wavelength-independent parameters are derived from the 68.3\% confidence region on each individual parameter using
\begin{equation}
\Delta\chi^{2} = \chi_i^{2} - \chi_{\rm min}^{2} = A\,,
\end{equation}
where $A$ depends on the number of wavelength-independent parameters (see Ref.\,\citenum{press07}).\\
Up to now 13 models can be fitted with GEM-FIND. The different models and their parameters are given in Table~\ref{modelparam}. The flow-chart in Fig.\,\ref{flow-chart} illustrates the fitting algorithm for GEM-FIND.


\section{Monte-Carlo simulations}
In order to test the reliability and limits of the geometrical models we derived synthetic visibilities with statistically applied errors for a subsample of the models. These synthetic visibilities were then used as input quantities for GEM-FIND, i.e.\,we simulated a statistically significant number of observations (sample size 50) and performed a least-squares analysis for each of them. In this work we will present Monte-Carlo simulations for the models  CircUD+CircGauss (M5), CircUD+EllGauss (M6) and UD+Dirac (M7) in Table\,\ref{modelparam}. Models M5 and M6 are optically thin and are especially suited to describe the circumstellar environment of e.g.\,asymptotic giant branch stars with low mass loss rate (e.g.\,Ref.\,\citenum{klotz12}). Model M7, on the other hand, is able to describe any system composed of a resolved primary star and an unresolved secondary component. In the following we will describe the idea and goals of the Monte-Carlo simulations and the determination of the synthetic visibilities. Results are discussed in Sect.\,\ref{results}.

\subsection{Idea and Goals}
\label{goals}
\begin{figure}
\includegraphics[bb=0 376 563 494,width=\textwidth]{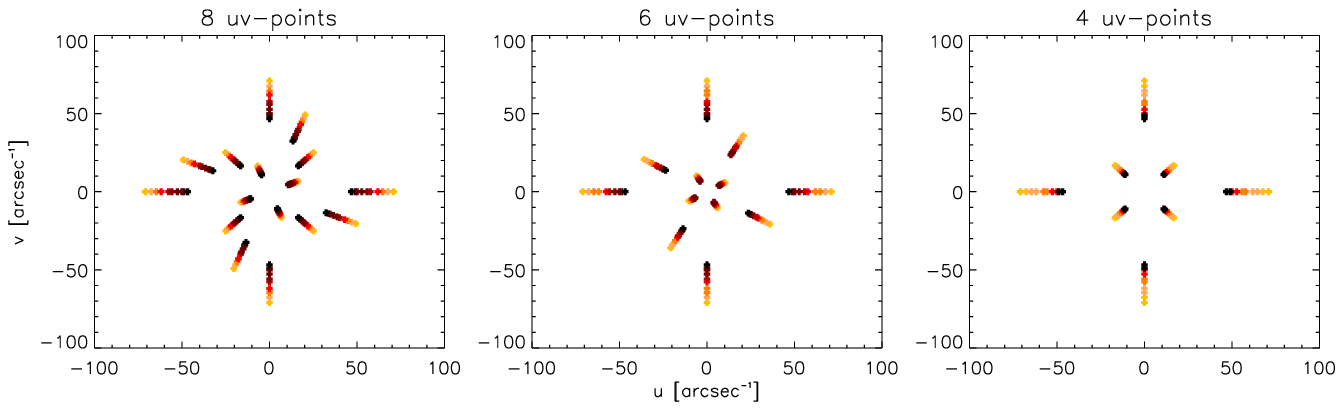}
\includegraphics[bb=24 376 563 494,width=16cm]{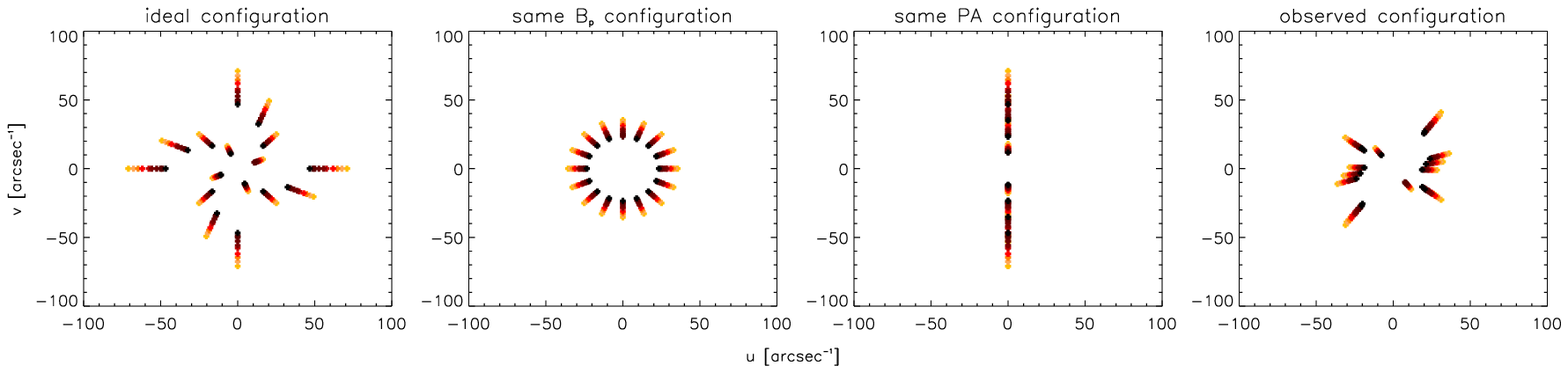}
\includegraphics[bb=0 376 563 494,width=\textwidth]{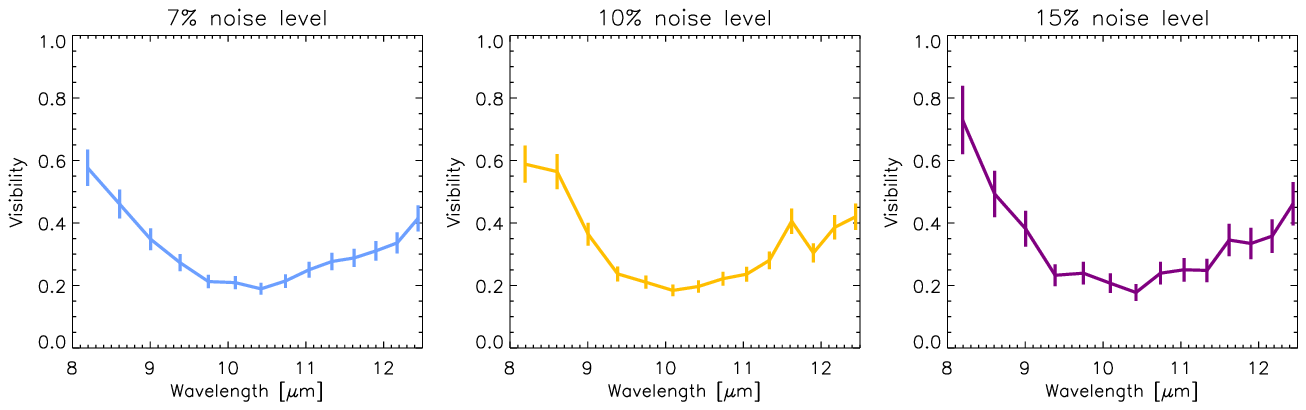}
\caption{\label{number-points}\label{configurations}\label{noise} \textbf{Upper row}: $uv$-coverage with $N_{uv}=8,6,4$ (from left to right) used to determine synthetic visibilities. Contour levels range from 8-12.5\,$\mu$m (light to dark, respectively) with a step size of 0.5\,$\mu$m. \textbf{Middle row}: $uv$-coverage of the configurations \textit{ideal}, \textit{sameB}, \textit{samePA}, \textit{obs} (from left to right) used to determine synthetic visibilities. \textbf{Lower row}: Examples of synthetic visibilities versus wavelength with different noise levels added (7, 10, 15\%; from left to right).}
\end{figure}

The Monte-Carlo simulations allow a numerical analysis of the stability of our fitting approach, thereby answering the following questions:
\begin{itemize}
\item \textbf{How stable is the parameter determination with respect to the number of $uv$-points $N_{uv}$?} \\
The focus is on a scarce $uv$-coverage as the limited number of telescopes (2 in the case of MIDI) often only provides a sparse sampling of the $uv$-plane. We determined synthetic visibilities with $N_{uv}=8, 6, 4$ (upper row in Fig.\,\ref{number-points}). 
\item \textbf{How does the distribution of $uv$-points affect the result?} \\
We selected 4 different baseline configurations (middle row in Fig.\,\ref{configurations}): (i) \textit{ideal} (baseline lengths B$_p$ and position angles PA are uniformly distributed in the $uv$-plane), (ii) \textit{sameB} (PA is uniformly distributed as in \textit{ideal}, B$_p$ remains unchanged), (iii) \textit{samePA} (B$_p$ is uniformly distributed as in \textit{ideal}, PA remains unchanged), (iv) \textit{obs} (we used the observed configuration from Ref.\,\citenum{klotz12}).
\item \textbf{How does observational noise influence the result?}\\
Different noise levels at FWHM $\sigma$ were applied to the synthetic visibilities. As shown in Ref.\,\citenum{chesneau07} the typical noise level of MIDI observations lies between 7\% (very good quality) and 15\% (acceptable quality). Therefore, we statistically applied Gaussian distributed errors of $\sigma=7,10,15$\% to the synthetic visibilities (lower row in Fig.\,\ref{noise}).
\item \textbf{How are the errors of the resulting parameters distributed?}
A least squares fitting minimization provides not only the best fitting values of the parameters but also the covariance matrix of the estimates that allows to calculate the uncertainties on the fitted parameters. These uncertainties are reliable if the errors on the parameters are known to be normally distributed. In order to guarantee this, observational errors have to be Gaussian distributed and the model must be linear. In certain cases even for a nonlinear model uncertainties can be derived from the covariance matrix (Ref.\,\citenum{lee10}), but this has to be tested using a Monte-Carlo simulation. If uncertainties are not Gaussian distributed, appropriate confidence regions have to be constructed. We determine the error of the results by calculating the difference between the true parameter value and the value calculated with GEM-FIND (summed over wavelength for wavelength-dependent parameters). To judge the level of normality a $\chi^2$ test is applied. This test provides the probability of getting the calculated $\chi^2$ for Gaussian distributed errors. Our null hypothesis is that errors are Gaussian distributed, i.e. a $\chi^2$ probability of 0.05 (this corresponds to a typical significance level of 5\%) or lower points to a statistically significant deviation from this hypothesis. The $\chi^2$ probability is given in the upper right corner in the insets of Fig.\,\ref{circud-circgauss-how-many-points}-\ref{ud+dirac-s40-which-error}.
\end{itemize}
Details on the synthetic visibility sets are given in Table\,\ref{tab:param}.


\subsection{Synthetic visibility determination}
With interferometry we observe the complex visibility of an object, i.e.\,the Fourier transform of its brightness distribution. The Earth's atmosphere influences the observations and makes it difficult to access the absolute phase of the object. Therefore, in the following only the absolute value of the complex visibility, the normalized visibility, will be considered.
Synthetic visibility functions can be derived from analytical formulae in the $uv$-plane, with $u=\frac{B_p}{\lambda}\sin{(\mathrm{PA})}$ and $v=\frac{B_p}{\lambda}\cos{(\mathrm{PA})}$. The synthetic visibility of model M5 can be derived by
\begin{equation}
V_\mathrm{CircUD+CircGauss}(u,v)=\left|\frac{f\frac{2 J_1(\pi r \theta_\mathrm{cen})}{\pi r \theta_\mathrm{cen}}+\exp\left(\frac{-(\pi r \mathrm{FWHM})^2}{4\ln 2}\right)}{f+1}\right|\;,
\end{equation}
with $J_1$ being the first order Bessel function of the first kind. The value of $r=\sqrt{u^2+v^2}$ depends on the configuration (\textit{ideal}, \textit{sameB}, \textit{samePA}, \textit{obs}) that is used.\\
Applying a rotation and compression to one axis (which becomes the semi-minor axis) of a circular model results in an ellipse, i.e.\, 
\begin{equation}
\mathrm{u}_\psi=\mathrm{u}\cos\psi - \mathrm{v}\sin\psi ~~\mbox{and}~~
\mathrm{v}_\psi=\mathrm{u}\sin\psi + \mathrm{v}\cos\psi\;,
\end{equation}
\begin{table}[h]
\caption{Details on the synthetic visibility sets. Given is the reference number of the synthetic visibility set, the number of $uv$-points, the type of configuration and the assumed noise level.} 
\label{tab:param}
\begin{footnotesize}
\begin{center}       
\begin{tabular}{cccc|cccc|cccc} 
\hline
\#$_\mathrm{synvis}$ & N$_{uv}$ & Config. & $\sigma$ [\%] & \#$_\mathrm{synvis}$ & N$_{uv}$ & Config. & $\sigma$ [\%] &\#$_\mathrm{synvis}$ & N$_{uv}$ & Config. & $\sigma$ [\%]  \\
\hline
\hline
S1 &  4     & ideal   & 7 &		S13 &  6     & ideal   & 7   &		S25 &  8     & ideal   & 7   \\
S2 &  4     & ideal   & 10 &	S14 &  6     & ideal   & 10   &		S26 &  8     & ideal   & 10   \\
S3 &  4     & ideal   & 15 &	S15 &  6     & ideal   & 15   &		S27 &  8     & ideal   & 15   \\
S4 &  4     & sameB  & 7  &		S16 &  6     & sameB  & 7   &		S28 &  8     & sameB  & 7   \\
S5 &  4     & sameB  & 10  &	S17 &  6     & sameB  & 10   &		S29 &  8     & sameB  & 10   \\
S6 &  4     & sameB  & 15  &	S18 &  6     & sameB  & 15   &		S30 &  8     & sameB  & 15   \\
S7 &  4     & samePA  & 7  &	S19 &  6     & samePA  & 7   &		S31 &  8     & samePA  & 7   \\
S8 &  4     & samePA  & 10  &	S20 &  6     & samePA  & 10   &		S32 &  8     & samePA  & 10   \\
S9 &  4     & samePA  & 15  &	S21 &  6     & samePA  & 15   &		S33 &  8     & samePA  & 15   \\
S10 &  4     & obs   & 7  &		S22 &  6     & obs   & 7   &		S34 &  8     & obs   & 7   \\
S11 &  4     & obs   & 10  &	S23 &  6     & obs   & 10   &		S35 &  8     & obs   & 10   \\
S12 &  4     & obs   & 15  &	S24 &  6     & obs   & 15   &		S36 &  8     & obs   & 15   \\
\hline\hline
\end{tabular}
\end{center}
\end{footnotesize}
\end{table} 

\begin{equation}
r_{\psi,\eta}=\sqrt{\mathrm{u}_\psi^2\eta^2+\mathrm{v}_\psi^2}\;,
\end{equation}
where $\psi$ denotes the inclination angle of the ellipse major axis (in sky plane, or minor axis in Fourier plane) measured from North to East and $\eta$ is the ratio of minor over major axis diameter $\eta=\theta_{min}/\theta_{maj}$. The visibility function for model M6 is then given by
\begin{equation}
V_\mathrm{CircUD+EllGauss}(u,v)=\left|\frac{f\frac{2 J_1(\pi r \theta_\mathrm{cen})}{\pi r \theta_\mathrm{cen}}+\exp\left(\frac{-(\pi r_{\psi,\eta} \mathrm{FWHM})^2}{4\ln 2}\right)}{f+1}\right|\;.
\end{equation}
The synthetic visibility of model M7 can be derived with
\begin{equation}
V_\mathrm{UD+Dirac}(u,v)=\left|\frac{\frac{2 J_1(\pi r \theta_\mathrm{prim})}{\pi r \theta_\mathrm{prim}}+f\exp(-2\pi i(u\Delta x+v\Delta y))}{1+f}\right|\;.
\end{equation}
A wavelength-independent, random relative error $\sigma_r$ is statistically applied to the synthetic visibilities by means of a random number generator and results in 
\begin{equation}
V_i^\mathrm{err}(u,v)=V_i(u,v)+\sigma_r\,,
\end{equation}
with $i$ defining the model that is used.
The probability distribution of the error is assumed to be normal with a FWHM of either 7\%, 10\% or 15\%. \\
The selection of the input parameters for $V_i$ will be discussed in the following. The parameters below are set in a way to describe a typical environment of an AGB star. The geometrical models implemented in GEM-FIND can be easily applied to other objects as well (e.g.\,YSO, AGNs).


\subsubsection{Choice of wavelength-independent parameters}
\underline{CircUD+CircGauss:} The diameter of the central star is fixed to $\theta_\mathrm{cen}=10$\,mas. The diameter of the central star is in practice often assumed to be known and therefore enters as a fixed parameter. \\
\begin{figure}
\centering
\includegraphics*[bb=54 376 473 490,width=15cm]{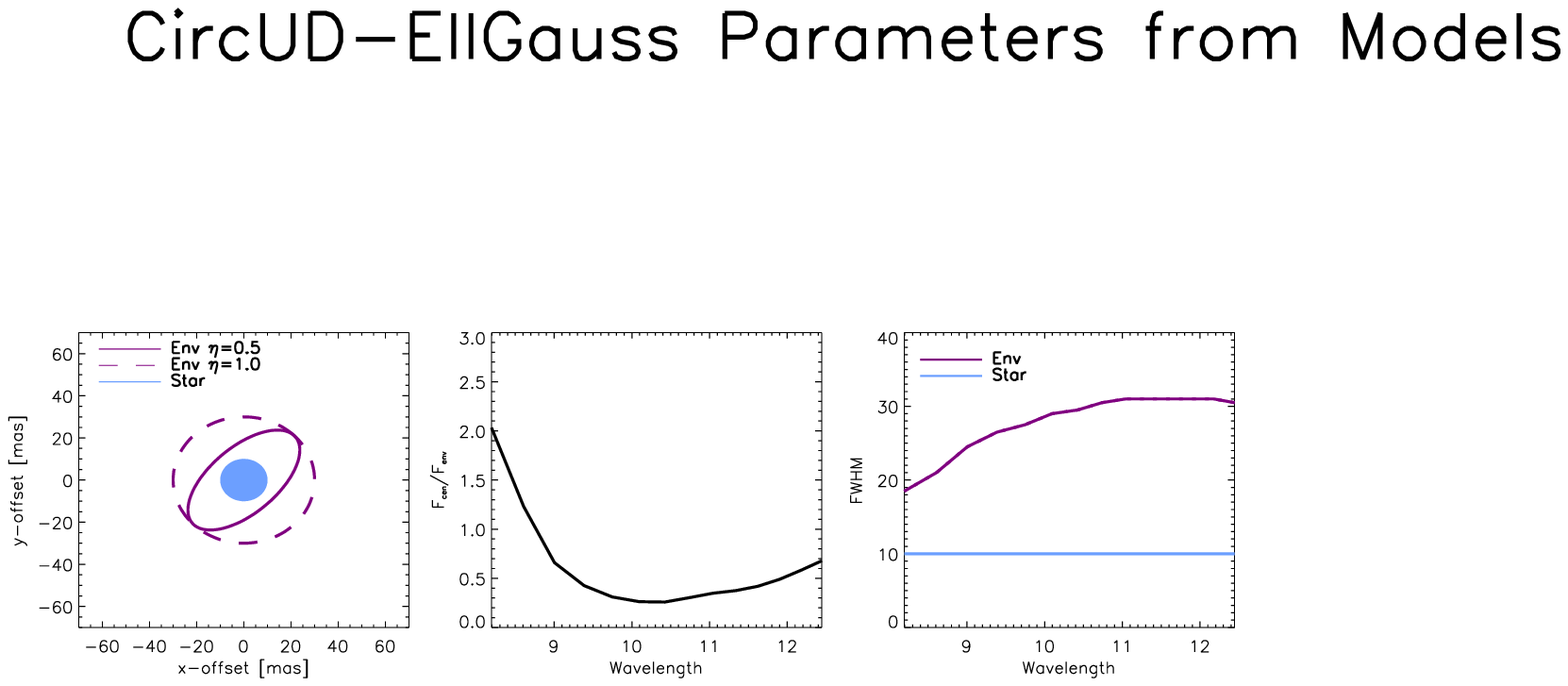}
\includegraphics*[bb=54 376 473 490,width=15cm]{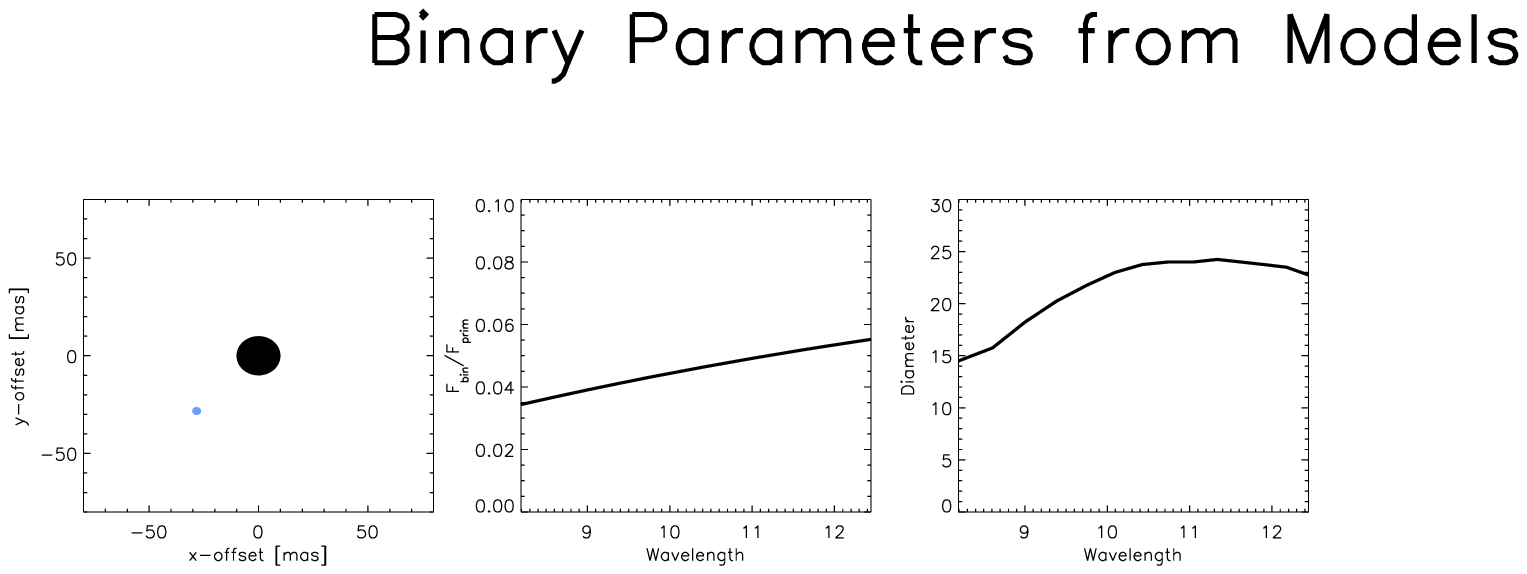}
\caption{\label{parameters-circud-circgauss} \label{parameters-circud-ellgauss} \label{parameters-ud+dirac}  \textbf{Upper row}: \label{parameters-circud-ellgauss} Parameters used for the synthetic visibility determination of CircUD+EllGauss and CircUD+CircGauss. Left: Wavelength-independent axis ratio $\eta$ for CircUD+EllGauss (full line) and CircUD+CircGauss (dashed line). Middle: Wavelength-dependent flux ratio $f$. Right: Wavelength-dependent FWHM of the envelope (dark violet line) as well as fixed diameter of central star (light blue line). \textbf{Lower row}: Parameters used for the synthetic visibility determination of UD+Dirac. Left: Wavelength-independent position of the unresolved companion. Middle: Flux ratio $f$ of the companion over the primary. Right: Wavelength-dependent diameter of the primary.}
\end{figure}

\underline{CircUD+EllGauss:} The diameter of the central star is fixed to $\theta_\mathrm{cen}=10$\,mas, the axis ratio $\eta=0.5$ and the position angle of the ellipse $\psi=45^\circ$.\\
\underline{UD+Dirac:} Observed separations of binary systems involving an AGB star lie between 7\,mas and 5'' (see Ref.\,\citenum{jorissen03}). In this work the separation $s$ of the companion (i.e. $s=\sqrt{\Delta x^2+\Delta y^2}$) is fixed to $s=40$\,mas.


\subsubsection{Choice of wavelength-dependent parameters}
\underline{CircUD+CircGauss:} The wavelength-dependent flux ratio $f$ as well as the FWHM of the Gaussian envelope were determined with the model of RR\,Aql (Fig.\,10 of Ref.\,\citenum{karovicova11}), a well studied oxygen-rich Mira variable. Ref.\,\citenum{karovicova11} studied the pulsation mechanism of the star by modeling multi-epoch interferometric observations with a radiative transfer code.
We set the flux ratio by dividing the flux of the central star by the flux of the envelope (upper panel in figure). The FWHM was set by fitting a CircUD+CircGauss model to the visibilities (lower panel in figure). The determined parameters are plotted in the upper row of Fig.\,\ref{parameters-circud-ellgauss}.\\
\underline{CircUD+EllGauss:} Parameters were set in the same way as described above. \\
\underline{UD+Dirac:} The diameter of the uniform disk was set as described above, but instead of a CircUD+CircGauss model a UD model was fitted to the visibilities (lower panel in Fig.\,10 of Ref.\,\citenum{karovicova11}). The flux ratio between the binary companion and the primary was set with the help of two blackbodies. One blackbody with a temperature of 400\,K (mean temperature of dust shell, Ref.\,\citenum{kerschbaum93}) was used to set the flux of the primary. A secondary component (e.g.\,planet) that is present in the surrounding of an AGB star will accrete matter from the mass that is lost by the star. Therefore, the other blackbody with a temperature of 300\,K (depending on the separation, Ref.\,\citenum{ivezic97}) was used to set the flux of the companion. Determined parameters are shown in lower row of Fig.\,\ref{parameters-circud-ellgauss}.

\section{Results}
\label{results}
The derived sets of synthetic visibilities (S1-S36, Table\,\ref{tab:param}) were used as input for GEM-FIND for models M5, M6 and M7. In the following we describe the result of the fit for each model separately, answering the questions raised in Sect.\,\ref{goals}.

\subsection{CircUD+CircGauss}
\label{CircUD+CircGauss}
\begin{figure}
\centering
\includegraphics*[bb=49 441 544 696,width=15cm]{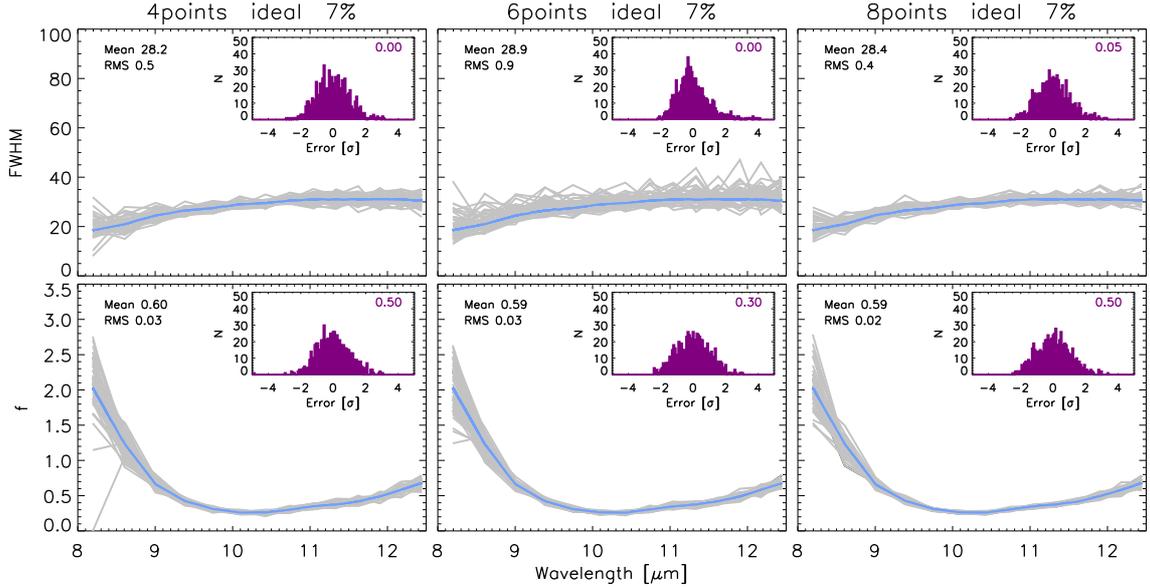}
\caption{\label{circud-circgauss-how-many-points} Parameters obtained with GEM-FIND for the synthetic visibility sets S1, S13, S25 (from left to right) of model CircUD+CircGauss: FWHM (upper panels) and flux ratio $f$ (lower panels) versus wavelength. The blue line corresponds to the true value. Wavelength-averaged means and standard deviations are denoted. The insets show the error distribution in the sample. The probability of normality is given in the upper right corner.}
\end{figure}
\begin{figure}
\centering
\includegraphics*[bb=49 431 544 700,width=15cm]{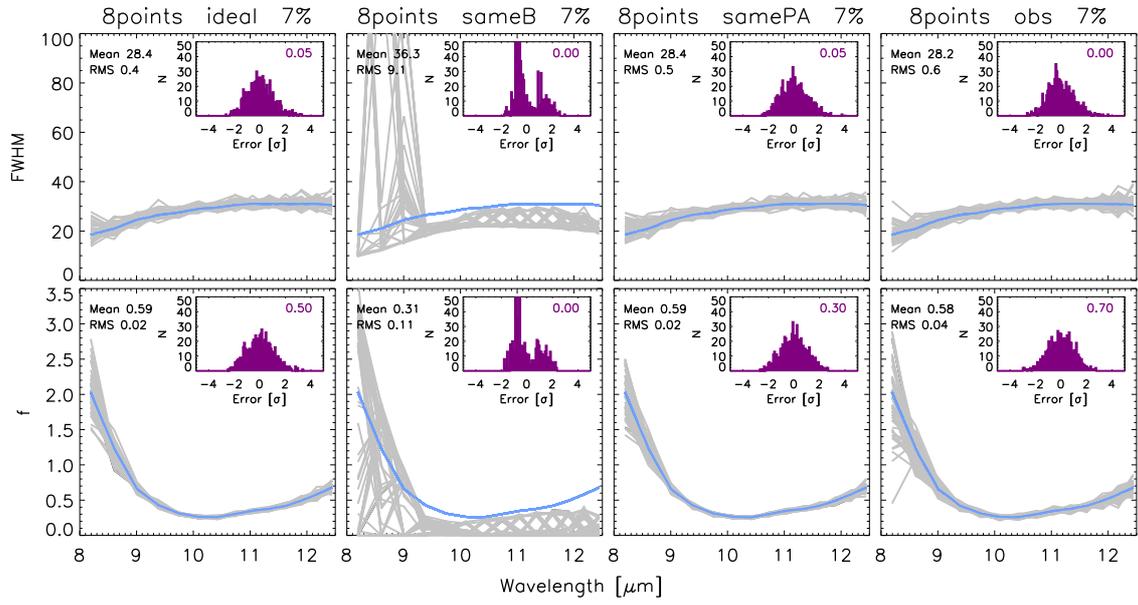}
\caption{\label{circud-circgauss-which-conf} Same as Fig.\,\ref{circud-circgauss-how-many-points} but for synthetic visibility sets S25, S28, S31 and S34 (from left to right) of model CircUD+CircGauss.}
\end{figure}

The CircUD+CircGauss model is spherically symmetric and has 3 input parameters (flux ratio $f$, FWHM of the Gaussian, central star diameter $\theta_{cen}$), where the first two are wavelength-dependent. The diameter of the central star is often assumed to be known and is therefore considered as a fixed wavelength-independent parameter, i.e. the effective number of free parameters $\mu$ is 2. 
\smallskip\\
\underline{How stable is the parameter determination with respect to the number of $uv$-points $N_{uv}$?} 
Fig.\,\ref{circud-circgauss-how-many-points} plots the results obtained with GEM-FIND for the synthetic visibility sets S1, S13, S25 (from left to right): FWHM (upper panels) and flux ratio $f$ (lower panels) versus wavelength. The insets show the error distribution in the sample between the true value and the values derived with GEM-FIND. The results for the flux ratio as well as the FWHM are comparable to the true value for $N_{uv}=4, 6, 8$. One would expect that the error distribution on the parameters is narrower the more $uv$-points are used. But, as can be seen from the figure, the RMS for 6 points is larger than the one for 4 points. This behavior is caused by the difference in the baseline lengths between the 4 points configuration (B$_p$=120, 120, 40, 40\,m) and the 6 points (B$_p$=120, 120, 70, 70, 20, 20\,m) configuration. The smallest baseline in the 6 points configuration is resolving the star only slightly.  The visibility of these points is very close to 1 and therefore these points do not add much information, i.e.\,we are left with 4 points also for the 6 points model. As the separation of the remaining baseline lengths is smaller than for the 4 points model, the determination of the right FWHM is more difficult.\\
To summarize, for this spherically symmetric model the parameter determination is stable for $N_{uv}=8, 6, 4$.
\smallskip\\
\underline{How does the distribution of $uv$-points affect the result?}
 Fig.\,\ref{circud-circgauss-which-conf}  plots the parameters obtained with GEM-FIND for the synthetic visibility sets S25, S28, S31 and S34 (from left to right). All configurations except \textit{sameB} (second column) find the right solution. \textit{sameB} uses the same baseline length for all points, i.e.\,in a spherical environment, all the points carry the same information (up to the synthetic errors). Therefore, the $uv$-coverage is too small to find the right solution. The similarity between the results of configurations \textit{ideal} (first column) and \textit{samePA} (third column) is expected for a spherically symmetric model, because the visibility profile looks the same no matter which position angle is used. The configuration \textit{obs} (fourth column) has a larger RMS for the flux ratio $f$, but is almost the same for the FWHM. This effect can be explained by the larger range in baseline length for configurations \textit{ideal} and \textit{samePA}, which is of greater importance for the determination of the flux ratio $f$. \\
In summary, the \textit{ideal} and \textit{samePA} are most suitable to find the right parameters for the spherically symmetric CircUD+CircGauss model. Also the \textit{obs} configuration is able to determine the correct parameters, but with a slightly broader error distribution. The \textit{sameB} configuration is not usable for spherically symmetric models with more than one component.
\begin{figure}
\centering
\includegraphics*[bb=49 441 544 696,width=15cm]{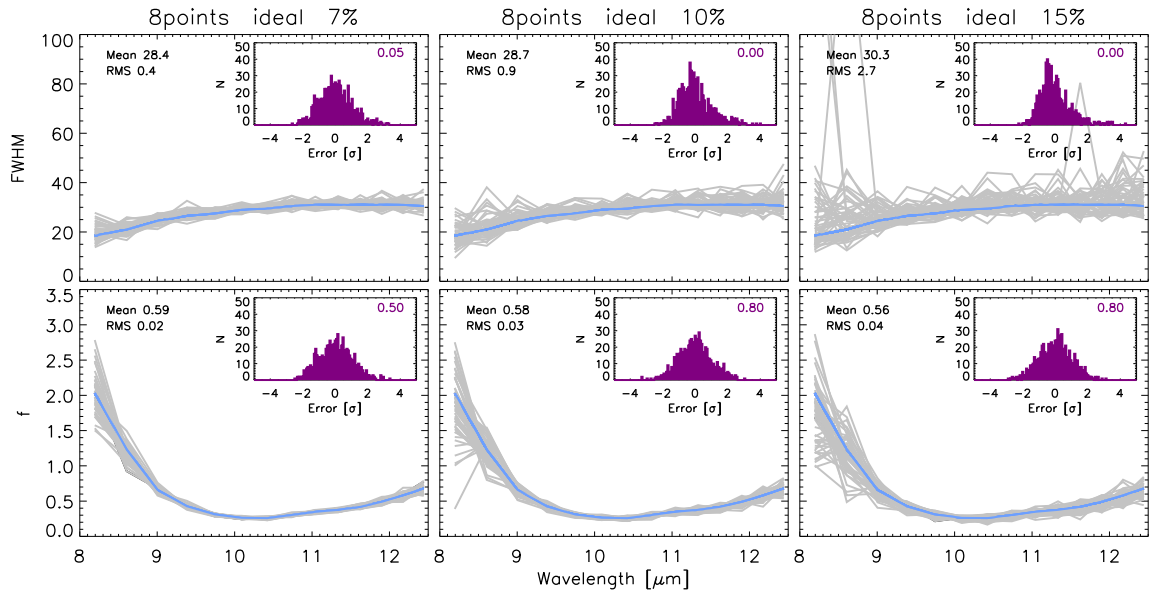}
\caption{\label{circud-circgauss-which-error}  Same as Fig.\,\ref{circud-circgauss-how-many-points} but for synthetic visibility sets S25, S26, S27 (from left to right) of model CircUD+CircGauss.} 
\end{figure}
\smallskip\\
\underline{How does observational noise influence the result?}
 Fig.\,\ref{circud-circgauss-which-error} plots the parameters obtained with GEM-FIND for the synthetic visibility sets S25, S26, S27 (from left to right). As expected, the larger the noise level on the visibilities the larger the RMS. The right solution is also found for a noise level as high as 15\%, even if the FWHM and flux ratio $f$ tend to be overestimated/underestimated, respectively. An even higher uncertainty may already be critical.
\smallskip\\
\underline{How are the errors of the resulting parameters distributed?}
The insets in Fig.\,\ref{circud-circgauss-how-many-points}, Fig.\,\ref{circud-circgauss-which-conf}  and Fig.\,\ref{circud-circgauss-which-error} show the error distribution in the sample (normalized and summed over all wavelengths) between the true value and the value calculated with GEM-FIND. The probability of normality is given in the upper right corner. The wavelength-averaged means and standard deviations given in the figures are very close to the true value FWHM$_\mathrm{true}$=28.0 and $f_\mathrm{true}$=0.60. The error distributions for the flux ratio are approximately Gaussian (probability $\geq$0.05) for all of the configurations except \textit{sameB}. On the other hand, error distributions for FWHM are not Gaussian distributed (probability $<$0.05). The FWHM tends to be slightly overestimated.\\

\subsection{CircUD+EllGauss}
\begin{figure}
\centering
\includegraphics*[bb=49 340 544 730,width=15cm]{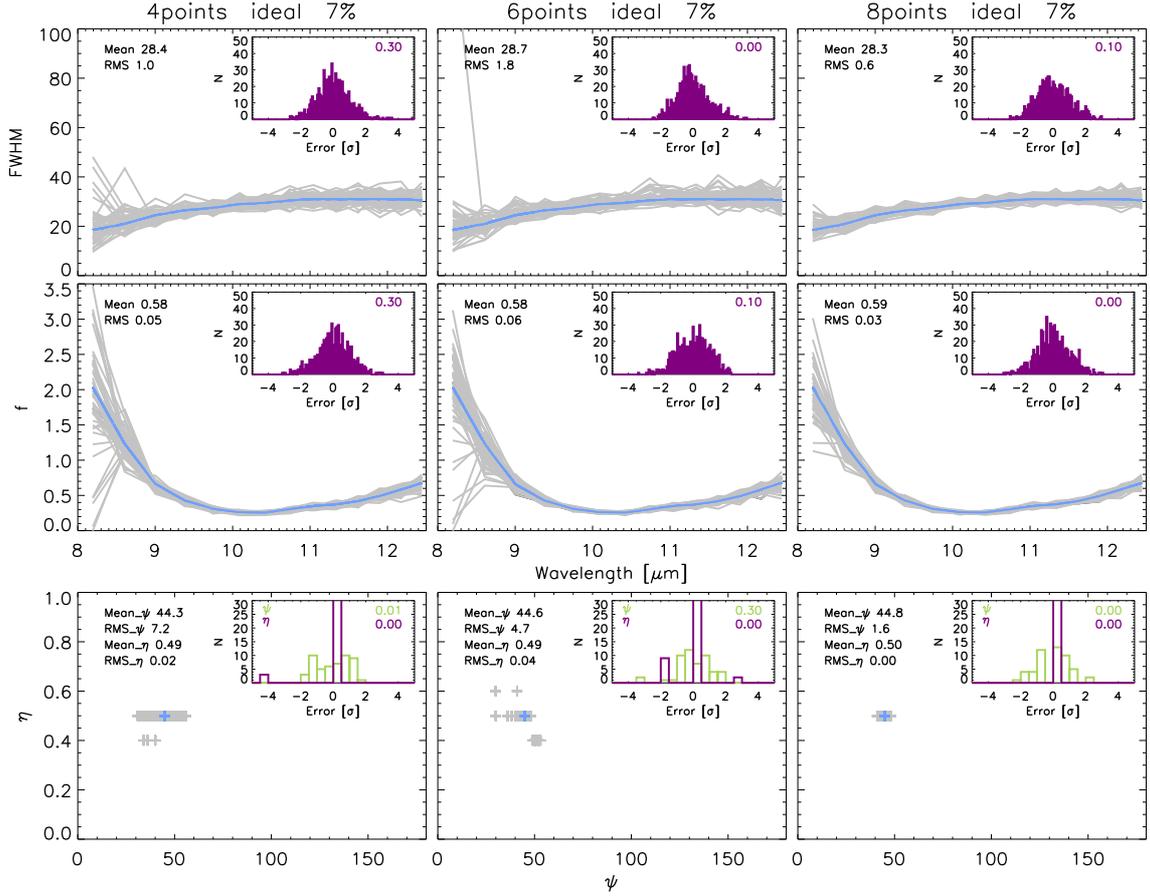}
\caption{\label{circud-ellgauss-eta050-how-many-points} Parameters obtained with GEM-FIND for the synthetic visibility sets S1, S13, S25 (from left to right) of model CircUD+EllGauss.   \textbf{Upper+middle}: FWHM (upper panels) and flux ratio $f$ (middle panels) versus wavelength. 
\textbf{Lower}: Axis ratio $\eta$ versus inclination angle $\psi$. The blue line/cross corresponds to the true value. Means and standard deviations are denoted. The insets show the error distribution in the sample. The probability of normality is given in the upper right corner.}
\end{figure}

\begin{figure}
\centering
\includegraphics*[bb=49 340 544 730,width=15cm]{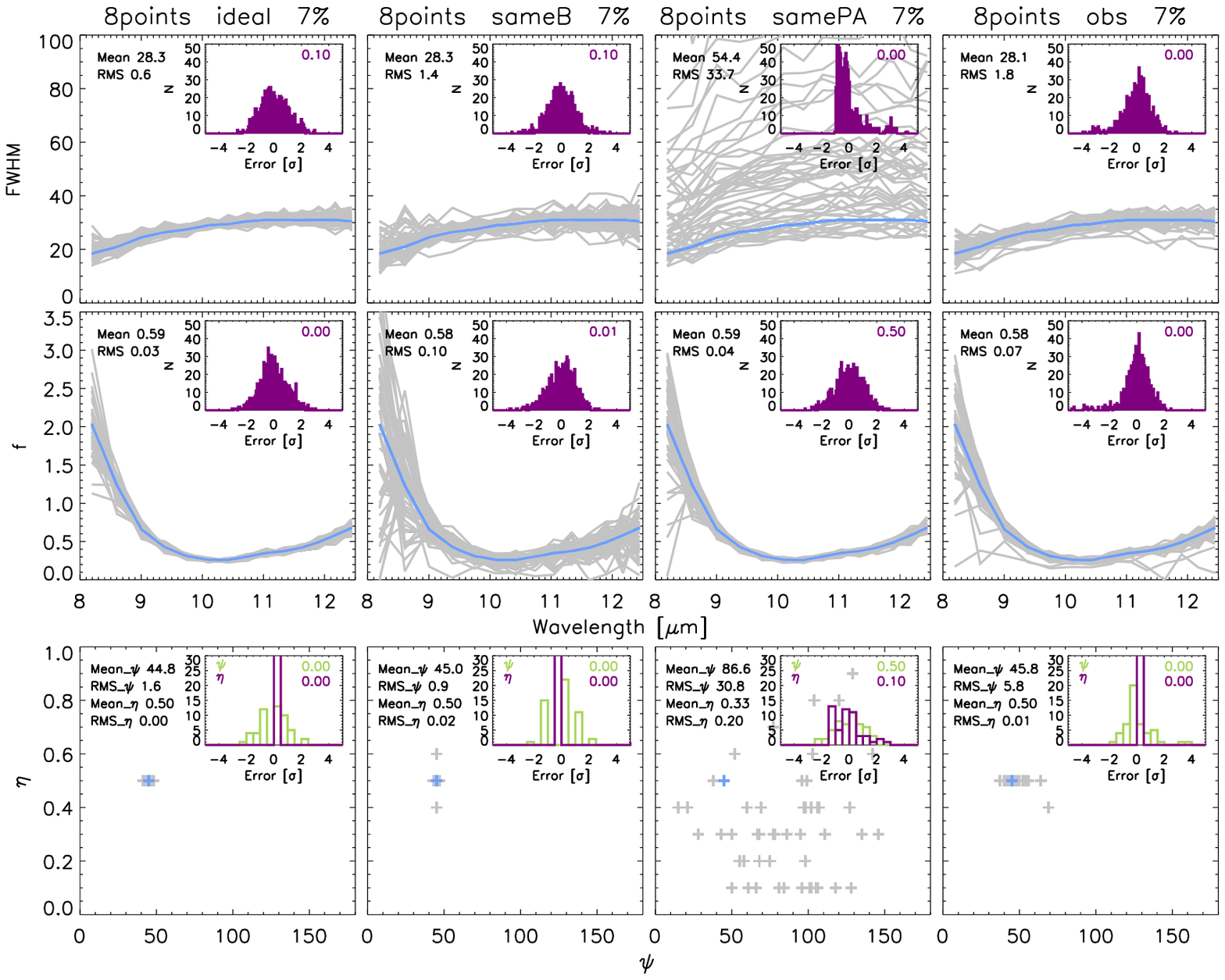}
\caption{\label{circud-ellgauss-eta050-which-conf} Same as Fig.\,\ref{circud-ellgauss-eta050-how-many-points} but for synthetic visibility sets S25, S28, S31 and S34 (from left to right) of model CircUD+EllGauss.}
\end{figure}

\begin{figure}
\centering
\includegraphics*[bb=49 340 544 730,width=15cm]{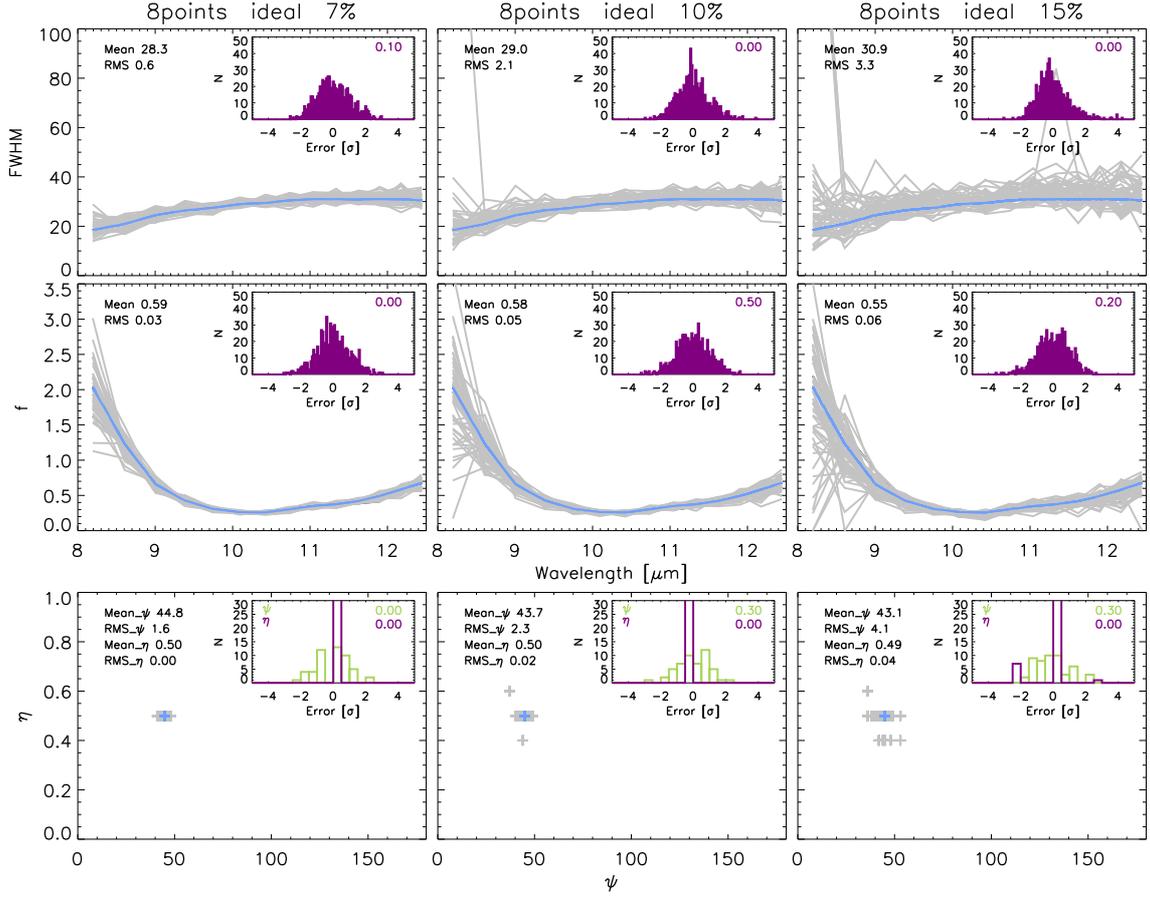}
\caption{\label{circud-ellgauss-eta050-which-error}  Same as Fig.\,\ref{circud-ellgauss-eta050-how-many-points} but for synthetic visibility sets S25, S26, S27 (from left to right) of model CircUD+EllGauss.} 
\end{figure}

The asymmetric CircUD+EllGauss model has 5 input parameters (flux ratio $f$, FWHM of the Gaussian, central star diameter $\theta_{cen}$, inclination angle of the ellipse $\psi$, axis ratio $\eta$), where the first two are wavelength-dependent. The diameter of the central star enters as a fixed parameter. 
\smallskip\\
\underline{How stable is the parameter determination with respect to the number of $uv$-points $N_{uv}$?}
Fig.\,\ref{circud-ellgauss-eta050-how-many-points} plots the parameters obtained with GEM-FIND for the synthetic visibility sets S1, S13, S25. For the FWHM (upper panels) and flux ratio $f$ (middle panels) the values found with GEM-FIND are very close to the true values for $N_{uv}=4, 6, 8$. As anticipated, the RMS is larger the less points are used. The observed behavior for 6 points described in \ref{CircUD+CircGauss} appears also in the case of an ellipse (evident from the mean and standard deviation values given in the figure). The correct solution for the axis ratio $\eta$ (y-axis in lower panels) is found for $N_{uv}=4, 6, 8$. On the other hand, the correct value for the inclination angle $\psi$ of the disk is only found in case of $N_{uv}=8$ and to a lesser extent also in case of $N_{uv}=6$.
\smallskip\\
\underline{How does the distribution of $uv$-points affect the result?}
Fig.\,\ref{circud-ellgauss-eta050-which-conf} plots the parameters obtained with GEM-FIND for the synthetic visibility sets S25, S28, S31 and S34. The \textit{ideal} (first column) and \textit{obs} (last column) configurations are both able to reliably determine the right value for all parameters. The configuration \textit{samePA} (third column) is not able to retrieve FWHM, $\eta$ and $\psi$, because using the same position angle for all points does not result in enough independent data points to determine the parameters of an ellipse. On the other hand, the flux ratio $f$ does not depend on the ellipticity (and consequently not on the position angle) and can therefore be determined by this configuration. The configuration \textit{sameB} (second column), on the other hand, is able to determine FWHM, $\eta$ and $\psi$. The flux ratio $f$ is more difficult to determine, because differences in flux ratio cause a more pronounced difference in the visibility profile at larger baselines. The larger baseline lengths ($\sim$120\,m) are not included in the configuration \textit{sameB} (largest baseline $\sim$60\,m), but covered by \textit{ideal} and \textit{samePA}. The largest baselines for \textit{obs} are $\sim$90\,m which accounts for the slightly larger RMS.
\smallskip\\
\underline{How does observational noise influence the result?}
Fig.\,\ref{circud-ellgauss-eta050-which-error} plots the parameters obtained with GEM-FIND for the synthetic visibility sets S25, S26, S27. The correct parameters are detected for 7\%, 10\% as well as 15\%. As expected, the larger the noise level on the synthetic visibilities the larger the RMS.
\smallskip\\
\underline{How are the errors of the resulting parameters distributed?}
The error distribution is most of the time non-Gaussian for all parameters (probability $<$0.05). This suggests that we cannot use the errors provided by the covariance matrix. Due to the coarse step size that is used for the axis ratio $\eta$ ($\Delta\eta=0.1$) it is not possible to determine the level of normality of the error distribution. Appropriate confidence regions should be constructed to determine the errors on all parameters.

\subsection{UD+Dirac}
\begin{figure}
\centering
\includegraphics*[bb=49 340 544 730,width=14.8cm]{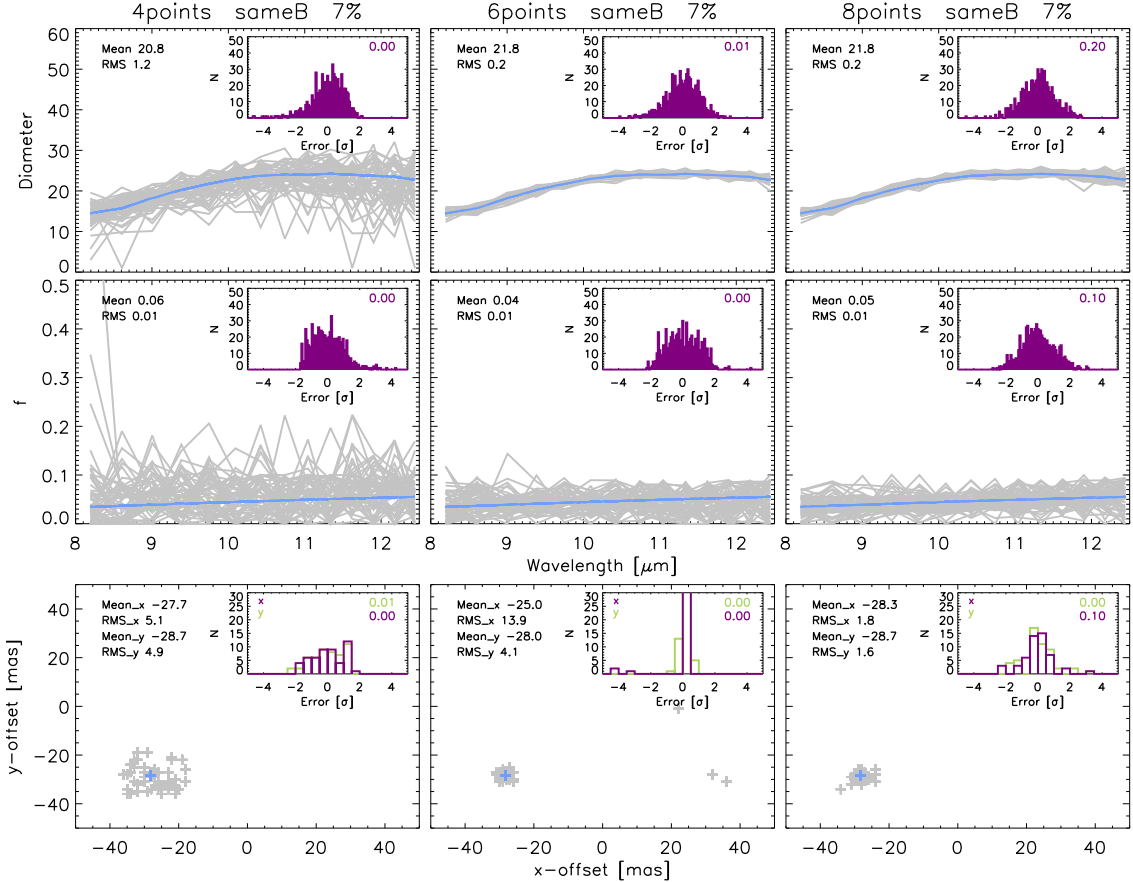}
\caption{\label{ud+dirac-s40-how-many-points} Parameters obtained with GEM-FIND for the synthetic visibility sets S4, S16, S28 (from left to right) of model UD+Dirac.   \textbf{Upper+middle}: UD diameter (upper panels) and flux ratio $f$ (middle panels) versus wavelength. 
\textbf{Lower}: x-position versus y-position of spot. The blue line/cross corresponds to the true value. The insets show the error distribution in the sample. The probability of normality is given in the upper right corner.}
\end{figure}

\begin{figure}
\centering
\includegraphics*[bb=49 340 544 730,width=15cm]{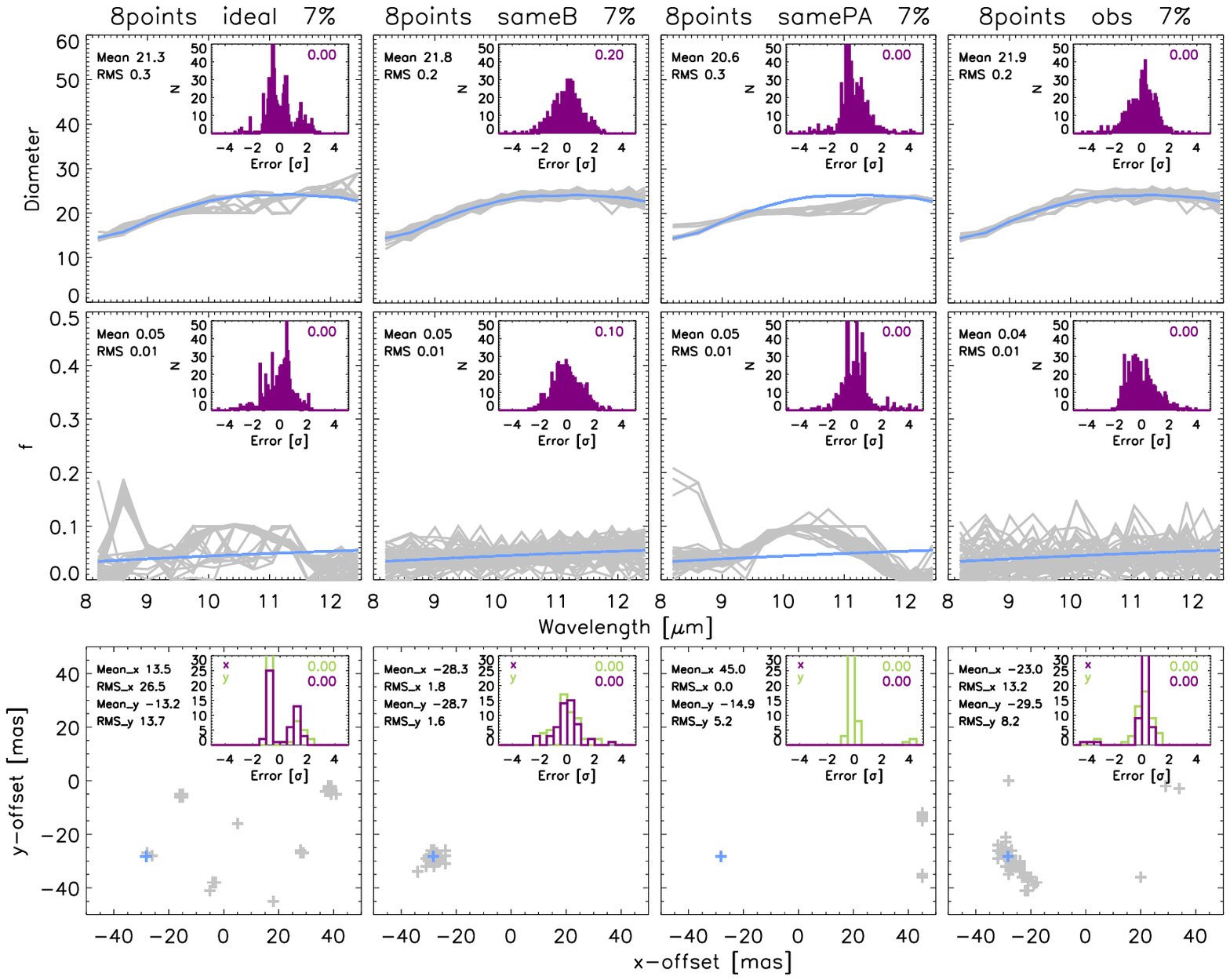}
\caption{\label{ud+dirac-s40-which-conf} Same as Fig.\,\ref{ud+dirac-s40-how-many-points} but for synthetic visibility sets S25, S28, S31 and S34 (from left to right) of model UD+Dirac.}
\end{figure}

\begin{figure}
\centering
\includegraphics*[bb=49 340 544 730,width=15cm]{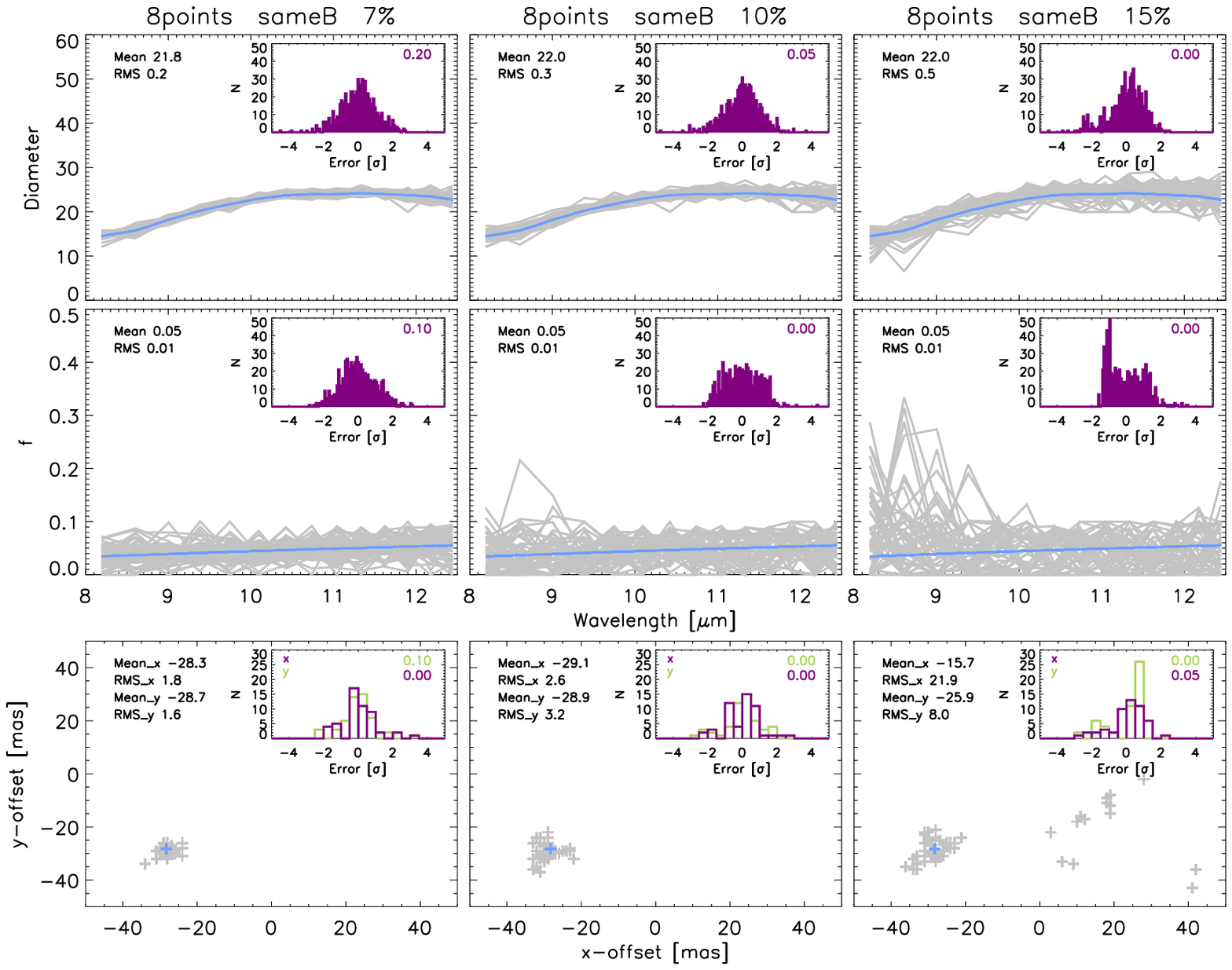}
\caption{\label{ud+dirac-s40-which-error} Same as Fig.\,\ref{ud+dirac-s40-how-many-points} but for synthetic visibility sets S28, S29, S30 from left to right) of model UD+Dirac.}
\end{figure}

The UD+Dirac model has 4 input parameters (position of the binary component $\Delta x$ and $\Delta y$, flux ratio $f$, diameter of the primary $\theta_\mathrm{prim}$), where the latter two are wavelength-dependent.\smallskip\\
\underline{How stable is the parameter determination with respect to the number of $uv$-points $N_{uv}$?}
Fig.\,\ref{ud+dirac-s40-how-many-points} plots the parameters obtained with GEM-FIND for the synthetic visibility sets S4, S16, S28. The correct values for the diameter and position of the companion can be found when 6 or more points are used. With 4 points  the parameter determination is no longer reliable (this is evident from the large RMS).  Note that the flux ratio can also be determined with 6 and 8 points, but with an uncertainty as large as $\pm25$\%. The flux ratio in our samples is quite small leading to a low contrast in visibilities. A larger flux ratio likely would result in a lower uncertainty. 
\noindent
\underline{How does the distribution of $uv$-points affect the result?}
Fig.\,\ref{ud+dirac-s40-which-conf} plots the parameters obtained with GEM-FIND for the synthetic visibility sets S25, S28, S31 and S34. The \textit{ideal} (first column) and \textit{samePA} (third column) configurations are not able to determine the expected values for the diameter, flux ratio and position of the companion. The reason for this behavior is the different baseline lengths, i.e.\,for the binary model it is advantegous to use baseline lengths as similar as possible. This statement is confirmed by the \textit{sameB} configuration and also by the \textit{obs} configuration that show only a small variation in baseline lengths. These configurations are able to find the right solution for all parameters. 
\smallskip\\
\underline{How does observational noise influence the result?}
Fig.\,\ref{ud+dirac-s40-which-error} plots the parameters obtained with GEM-FIND for the synthetic visibility sets S28, S29, S30. Noise levels of 7\% and 10\% are able to determine the right parameters. Also a noise level of 15\% is still able to determine the diameter and position of the companion with a reasonable error distribution. On the other hand, the determination of the flux ratio is no longer possible with such a high noise level. As mentioned before, we are confident that a larger value of the flux ratio would allow to find the right solution with a lower uncertainty.
\smallskip\\
\underline{How are the errors of the resulting parameters distributed?}
The error distribution of the parameters is non-Gaussian for most synthetic visibility sets. Also in this case it is recommended to use confidence regions for the error estimation. 

\newpage
\section{Conclusions}
In this work we presented the geometrical model fitting tool GEM-FIND for MIDI interferometric data. In order to guarantee reliability and to test the limits of the fitting method a Monte-Carlo approach was used to generate synthetic visibilities with statistical noise of three models (CircUD+CircGauss, CircUD+EllGauss, UD+Dirac). We used synthetic data sets with different $uv$-coverage, number of $uv$-points and noise levels in order to be able to answer the following questions: How many $uv$-points are needed for a correct parameter determination? How do these points have to be distributed in the $uv$-plane? How large can the noise level of the observations be?\\
The following main results were found: As expected, it is recommended to (i) avoid the use of similar baseline lenghts (sameB) to determine the parameters of a spherical object morphology; (ii) avoid the use of similar position angles (samePA) to determine the parameters of an elliptical object morphology. (iii) We find that the use of close baseline lengths (\textit{sameB} and \textit{obs}) facilitates the determination of the parameters of a binary system. Here great care must be taken when the morphology of an object with unknown geometry is to be determined, i.e. one has to account for the effect of distribution of $uv$-points on the fitting results. 
(iv) Our study shows that observations with a noise level as high is 15\% can still be used to determine the parameters in most of the cases. (v) As the error distribution of the output deviates from normality, errors of all parameters should be derived using confidence regions.

\acknowledgments         
 
This work is supported by the Austrian Science Fund FWF under project number AP23006. 


\bibliography{gemfind}   
\bibliographystyle{spiebib}   

\end{document}